\pdfoutput=1
\documentclass{IEEEtran}

\usepackage{graphicx}
\usepackage{amsmath,amssymb}
\usepackage{siunitx}
\usepackage{booktabs}
\usepackage{multirow}
\usepackage{cite}
\usepackage{url}
\usepackage{xcolor}
\usepackage{soul}
\usepackage{xcolor}
\sethlcolor{yellow}
\usepackage{booktabs,tabularx}
\usepackage[hidelinks]{hyperref}
\usepackage{verbatim}
\usepackage{algorithm}
\usepackage{algpseudocode}
\algrenewcommand{\algorithmicrequire}{\textbf{Input:}}

\begin{document}
\title{Deep Reinforcement Learning Assistance for Resilient Haptic Teleoperation over a 5G Network}

\author{Georgios~Kokkinis,~\IEEEmembership{Graduate~Student~Member,~IEEE},
        Aitor~Hernandez~Herranz,
        and~Qi~Zhang,~\IEEEmembership{Senior~Member,~IEEE}%
\thanks{G. Kokkinis and Q. Zhang are with DIGIT and Department of Electrical and Computer Engineering, Aarhus University, Aarhus, Denmark (e-mail: \{gkokkinis, qz\}@ece.au.dk).}%
\thanks{A. Hernandez Herranz is with Ericsson Research, Stockholm, Sweden (e-mail: aitor.hernandez.herranz@ericsson.com).}%
\thanks{This research was supported by the TOAST project, funded by the European Union's Horizon Europe research and innovation program under the Marie Sk{\l}odowska-Curie Actions Doctoral Network (Grant Agreement No. 101073465), the Danish Council for Independent Research project eTouch (Grant No. 1127-00339B), and NordForsk Nordic University Cooperation on Edge Intelligence (Grant No. 168043).}%
\thanks{(Corresponding authors: Georgios Kokkinis; Qi Zhang.)}%
}

\markboth{IEEE Transactions on Mobile Computing,~Vol.~XX, No.~X, Month~2026}%
{Kokkinis \MakeLowercase{\textit{et al.}}: DRL Assistance for Resilient Haptic Teleoperation over 5G}

\maketitle
\begin{abstract}
Haptic teleoperation requires low-latency and reliable communication to preserve stable and responsive interaction between a human operator and a remote environment. However, network congestion can strongly degrade teleoperation performance, particularly when haptic traffic competes with other data flows for limited radio resources. This work investigates a deep reinforcement learning (DRL)-assisted teleoperation framework over an Ericsson 5G standalone (SA) testbed under different congestion and quality-of-service (QoS) configurations. In the evaluated task, a human operator uses a haptic device to steer a spaceship through a remote virtual game. The DRL agent runs inside the remote environment, hosted at the network edge, and assists the operator by blending its own actions with the operator's commands received over the 5G network. This placement gives the agent direct access to the current state of the environment, unaffected by network delay. Experimental results show that DRL assistance improves task performance under all network conditions, with the largest benefit under congestion, where it reduces mean task completion time by 20.6\%. Under congestion, assisted operation even outperforms unassisted operation with QoS prioritisation, which restores latency close to the uncongested baseline.
\end{abstract}


\begin{figure*}[t]
    \centering
    \includegraphics[width=\textwidth]{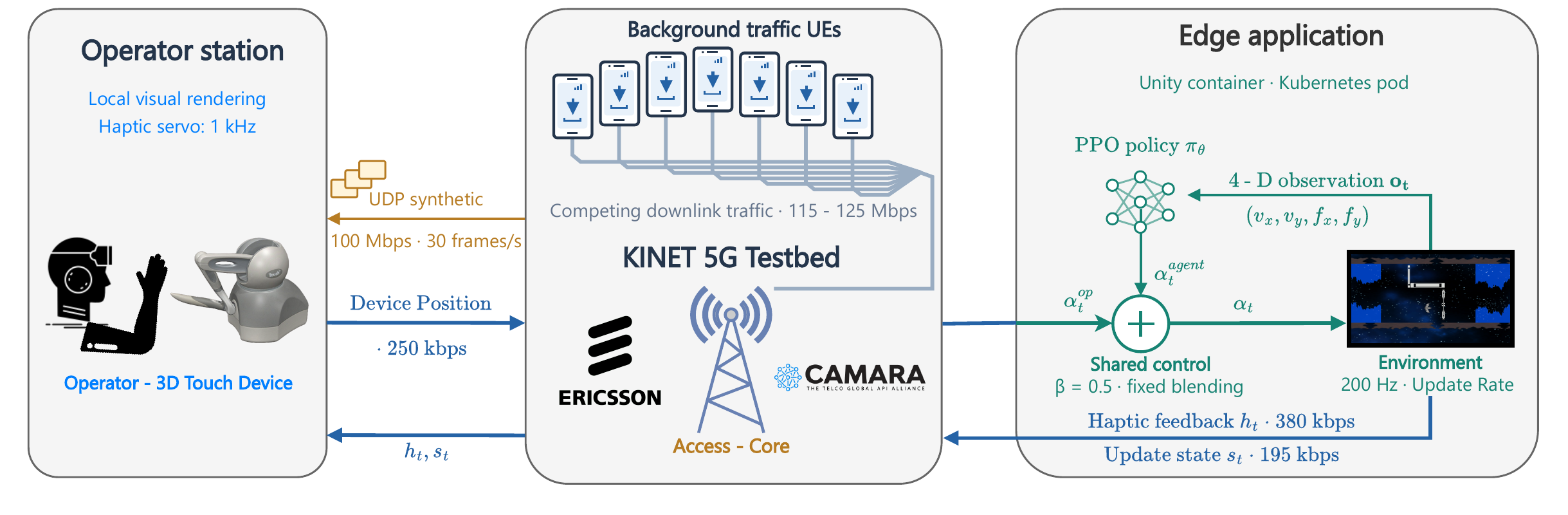}
    \caption{System architecture of the DRL-assisted haptic teleoperation platform. Operator commands traverse the 5G network and are blended with the DRL policy's actions within the edge-hosted Unity application. The policy accesses the environment state directly, while state and haptic feedback return to the operator station over the network.}
    \label{fig:system-architecture}
\end{figure*}

\section{Introduction}
\label{sec:introduction}

The Tactile Internet extends network-mediated teleoperation beyond audio and video by allowing a human operator to interact physically with a remote environment~\cite{fettweis2014tactile}. The human operator sends control commands and receives haptic feedback, forming a closed loop over the network~\cite{steinbach2012haptic}. This makes teleoperation highly sensitive not only to average latency, but also to how latency evolves during the task~\cite{aijaz2017realizing}. A delayed or inconsistent response can reduce task precision, degrade the human operator's control transparency, and eventually affect the stability of the interaction.

This becomes particularly challenging over a shared mobile network, where delay is not constant. Wireless channel conditions vary over time, users compete for resources, and sustained traffic can create queues that grow during operation. Studies of teleoperation under delay often isolate the control problem by injecting fixed or stochastic latency into an otherwise ideal connection~\cite{gulecyuz2025enhancing}. This is useful for evaluating the effect of delay, but it treats the network as an external parameter. In a typical cellular system, the evolution of delay during task execution exposes the human operator to changing communication conditions. Measuring network behaviour alongside task outcomes is therefore necessary to examine how these variations relate to teleoperation performance.

Existing work addresses this problem from three directions. On the network side, quality-of-service (QoS) mechanisms and network slicing prioritise latency-sensitive traffic, shaping the delivered latency and throughput~\cite{hamidovic2023campus}, and learned resource allocation can adapt this prioritisation dynamically~\cite{golmohammadi2025hierarchical,kokkinis2025slicing}. At the signal level, learning-based haptic signal prediction estimates delayed or lost haptic samples at the receiver, relaxing the delay bound that the network must satisfy~\cite{kokkinis2025delaybound,kokkinis2026xhap}, and feature-attribution methods improve the robustness of such predictors~\cite{vahedifar2026dmd,vahedifar2026shapley}. At the endpoint, shared autonomy blends the operator's command with an autonomous policy to support them when the control loop degrades~\cite{dragan2013policy,javdani2018shared,reddy2018shared}, including in teleoperation under delay~\cite{gulecyuz2025enhancing,abouallaban2019blended}.

These strategies are well established but usually studied in isolation. Network studies characterise congestion and QoS through network metrics such as latency, jitter, and throughput, without linking them to task-level metrics such as completion time and accuracy. Signal prediction is typically evaluated through prediction error on recorded haptic traces, and learning-based assistance under injected delay only. Congestion in real networks, however, produces latency that is uncertain and fluctuates during the task. It therefore remains unclear whether assistance still benefits the human operator over a live network, and how it interacts with QoS prioritisation, in which the network schedules the teleoperation traffic ahead of competing traffic to preserve its latency.

In this paper, we present an assistive deep reinforcement learning (DRL) agent for haptic teleoperation and evaluate its effect on task performance over a live Ericsson 5G standalone (SA) testbed. The task is a virtual navigation game in which the human operator steers a spaceship with a haptic device while avoiding a penalty region, the water-zone. The water-zone is invisible to the operator and is perceived only through haptic feedback. Its high viscosity reduces the spaceship's maximum speed and produces a drag force that opposes motion, making the spaceship slower and harder to steer, while a restoring force pushes the spaceship towards the empty space, outside of the water-zone. The agent is pretrained offline in the same virtual environment, with haptic feedback as its observation, so that it responds to the same contact forces the human operator perceives.  

As shown in Fig.~\ref{fig:system-architecture}, the agent runs inside the remote environment and reads its state directly, so it continues to act on up-to-date information even when the human operator's commands and state updates are delayed by the network. The agent can therefore compensate when the operator's control loop degrades. To the best of our knowledge, this is the first evaluation of a DRL-based haptic teleoperation assistant over a live 5G SA network under real congestion, rather than an emulated environment.

We evaluate the system under three network conditions: an uncongested baseline, a congested cell, and a congested cell in which the teleoperation traffic receives QoS priority~\cite{3gpp23501}. During each session, we jointly record radio conditions, radio access network (RAN) Cell Trace Record (CTR) data, active transport-layer latency probes, and application-layer task outcomes. The CTR data provide high-resolution radio-layer events collected directly from the 5G testbed. The QoS configuration and network infrastructure use Ericsson production components, allowing us to evaluate their interaction with autonomous assistance and task performance on a real 5G network.

The main contributions of this work are summarised as follows:

\begin{itemize}

\item \textbf{Haptic-aware DRL assistant for teleoperation:}
We develop a DRL assistant, pretrained offline with haptic feedback in its observation space, that blends its actions with the human operator's commands. By executing alongside the virtual environment and acting directly on its current state, the assistant remains effective when network delay degrades the human operator's control loop.

\item \textbf{Evaluation of DRL assistance on a live 5G network:}
We evaluate the assistant over a live Ericsson 5G SA testbed under uncongested, congested, and QoS-prioritised conditions. For the evaluated human operator, assistance reduces mean task completion time by 15.5--20.6\% and mean water-zone time by approximately 40\% compared with human-only operation. The largest completion-time reduction occurs under congestion, showing that the assistant is most valuable during network instability. The system setup and demo application will be made publicly available upon publication.

\item \textbf{Cross-layer characterisation of the network conditions:}
By jointly measuring the radio, RAN, transport, and application layers in every session, we show that congestion not only raises latency but makes it grow while the task runs, by 5.5\,ms/s on average, before partially recovering between sessions. This couples task duration to the latency the operator experiences and explains why DRL assistance, which shortens the task, is most beneficial under congestion. We further demonstrate that QoS prioritisation removes this accumulation and restores latency close to uncongested conditions at the same cell load, reducing the median round-trip time (RTT) from 113.2\,ms to 15.7\,ms.
\end{itemize}

The remainder of this paper is organised as follows. Section~\ref{sec:related} reviews related work on shared autonomy and network-supported teleoperation. Section~\ref{sec:system} describes the system architecture and the DRL assistant, and Section~\ref{sec:testbed} introduces the 5G testbed and measurement setup. Section~\ref{sec:method} presents the experimental design and evaluation metrics. Section~\ref{sec:results} reports the effects of congestion, DRL assistance, and QoS prioritisation. Section~\ref{sec:discussion} discusses latency accumulation and the limitations of the study. Finally, Section~\ref{sec:conclusion} concludes the paper.

\section{Related Work}
\label{sec:related}

Shared autonomy combines the human operator's input with actions proposed by an autonomous policy. The policy-blending framework in~\cite{dragan2013policy} predicts the operator's goal and arbitrates between human and autonomous commands, while the approach in~\cite{javdani2018shared} accounts for uncertainty over possible goals. DRL was subsequently applied to shared autonomy in~\cite{reddy2018shared}, using online learning and value-based action filtering. These methods assume that the operator's commands reach the policy without delay, an assumption that no longer holds when teleoperation runs over a network.

Several studies have therefore examined shared autonomy under communication delay. The approach in~\cite{gulecyuz2025enhancing} learns from teleoperated demonstrations and adapts the blending of human and autonomous commands according to communication delay and confidence in the learned autonomy, while~\cite{abouallaban2019blended} considers blended assistance that addresses operator input latency and erroneous odometry. In both cases, delay is introduced artificially rather than arising from a real network. Our system follows the same shared-control principle, combining a policy trained in simulation with the operator's commands through fixed-weight blending, and evaluates it under the delay produced by a live, congested 5G network.

Model-mediated teleoperation (MMT) addresses communication delay by rendering feedback at the operator side from a local model of the remote environment, updated with remote measurements~\cite{mitra2008model}. Its performance depends on how closely this model tracks the remote environment, since model mismatch distorts the rendered feedback and network-delivered corrections can cause abrupt force changes. The method in~\cite{Mads2024} improves MMT through an octree-based environment representation, while~\cite{kroep2025intent} incorporates operator intent to support task execution when the local model and the remote environment diverge. A related implementation combines predictive haptic feedback with live video and characterises latency over a private 5G network~\cite{Kroep2025BreakingTL}. Our system also maintains an environment replica at the operator station, but uses it only for visual rendering: the haptic device renders the most recent force received from the remote environment, and assistance is applied at the remote endpoint by blending the operator's commands with an autonomous policy.

Haptic signal prediction compensates for delayed or lost feedback by estimating the missing haptic samples at the receiver. The deep-learning-based estimator in~\cite{kokkinis2025delaybound} bounds the estimation error so that the delay requirement of haptic packets can be relaxed, allowing multiple packets to be batched into a single resource block and increasing the number of reliably served users. The cross-modal attention architecture in~\cite{kokkinis2026xhap} fuses the history of force feedback with the operator's control actions, reducing the prediction error and lowering the signal-to-noise ratio required for reliable transmission. Robustness of such predictors is improved in~\cite{vahedifar2026shapley} through Shapley-value-based feature selection and in~\cite{vahedifar2026dmd} by combining discrete mode decomposition with Shapley mode values. These methods improve the feedback presented to the operator, which is complementary to assisting the commands sent to the environment, as done in this work.

The placement of control and computation also shapes networked teleoperation. The system in~\cite{ahmed2025splitcontrol} uses 5G mobile edge computing for edge-driven split control, applying edge-based prediction and control to assist human operation under communication delay. Its experiments include a seesaw-like robot and a haptic-glove-controlled robotic hand. Communication--compute co-design is investigated in~\cite{Fresa2025}, where reinforcement learning distributes computation across the device--edge--cloud continuum of a mobile robot application under network service-level agreements. In our system, the assistive policy is embedded in the virtual environment application, giving it direct access to the simulation state.

Network-focused approaches improve the communication conditions supporting teleoperation. Measurements in~\cite{hamidovic2023campus} characterise how QoS configuration affects throughput and latency in a loaded 5G campus network. DRL has been used to allocate resources between enhanced mobile broadband (eMBB) and ultra-reliable low-latency communication (URLLC) queues under application delay constraints~\cite{golmohammadi2025hierarchical} and to slice radio resources between the video and haptic modalities of teleoperation~\cite{kokkinis2025slicing}. At the application endpoint, the framework in~\cite{islam2022edge} combines congestion prediction with Q-learning to select haptic codecs, evaluated on recorded traces over an emulated network. These approaches act on resource allocation or traffic generation, whereas our assistive policy acts on the command applied to the environment, and our experiments combine it with network QoS prioritisation.

Studies on active queue management~\cite{nichols2012controlling} and buffer sizing~\cite{spang2021updating} show that sustained traffic load fills network buffers, causing delay to grow over time. Since a session-level average does not capture such growth, we examine within-session latency changes and session-boundary behaviour alongside task performance.

Unlike these works, which address command-level assistance under injected delay, signal-level prediction of haptic feedback, or network-level mechanisms without task-level evaluation, we combine DRL assistance with QoS prioritisation on a live 5G network and relate the latency experienced by the operator to task performance.

\section{System Architecture and DRL Assistant}
\label{sec:system}

This section describes the DRL-assisted haptic teleoperation system. We first present the teleoperation task and the haptic loop, followed by the split between execution of the remote environment at the network edge and rendering at the operator station. We then describe the assistive agent, including its observations, shared-control formulation, and training procedure. The 5G testbed and measurement setup are described separately in Section~\ref{sec:testbed}. The notation used throughout the paper is summarised in Table~\ref{tab:notation}.

Fig.~\ref{fig:system-architecture} gives an overview of the system. At the human operator station, the position of the haptic device is mapped to an operator command, which is transmitted over the uplink of the KINET 5G testbed to the edge application. There, the DRL assistant computes its own command from the environment state, the two commands are blended (Section~\ref{sec:agent}), and the result is applied to the virtual game, i.e. the remote environment. The updated state and haptic feedback from the remote environment return to the operator station over the downlink and are rendered locally (Section~\ref{sec:architecture}). The assistant therefore acts on current state, whereas the operator acts based on the state that has traversed the network. Fig.~\ref{fig:system-architecture} also shows the competing traffic used to load the network: in the congested and prioritised conditions, background UEs and a synthetic video-like stream share the cell with the teleoperation traffic, i.e., the operator commands on the uplink and the state and haptic feedback on the downlink, so that the system can be evaluated under congestion (Section~\ref{sec:traffic}).

\subsection{Teleoperation Task and Haptic Loop}
\label{sec:task}

The human operator uses a 3D Systems Touch~\cite{3dsystems_touch} haptic device to guide a spaceship vehicle through a fixed
course. The course contains gate obstacles that the human operator must navigate and an invisible water region indicated through haptic cues. Water contact reduces vehicle speed and is penalised. The goal of the task is for the spaceship to cross the finish line within the course. The haptic device provides three degrees of freedom, supports forces up to 3~N, and supports force updates at rates up to 1~kHz. Contact with obstacles and environmental forces are rendered back to the human operator's hand. The task is two-dimensional: the spaceship moves in a vertical plane, with $x$ denoting the forward (horizontal) direction and $y$ the vertical direction. Although the haptic device provides three degrees of freedom, only its two in-plane axes are used for commands and force feedback.

\begin{table}[t]
\centering
\caption{Notation}
\label{tab:notation}
\small
\setlength{\tabcolsep}{2pt}
\renewcommand{\arraystretch}{1.0}
\begin{tabularx}{\columnwidth}{@{}lXlX@{}}
\toprule
\textbf{Symbol} & \textbf{Definition} &
\textbf{Symbol} & \textbf{Definition} \\
\midrule
$t$ & Physics step &
$\rho$ & Normalised offset \\
$k$ & Decision step &
$y_{\mathrm{ship}},y_c$ & Ship / centre height \\
$\delta t$ & Physics timestep &
$w$ & Channel height \\
$\Delta t_k$ & Decision interval &
$r_k$ & Reward \\
$D$ & Decision period &
$r^{\mathrm{term}}_k$ & Terminal reward \\
$\mathbf{p}^{\mathrm{op}}_t$ & Device position &
$\kappa_{(\cdot)},R_{(\cdot)}$ & Reward weights \\
$\mathcal{M}$ & Command mapping &
$\Delta x_k$ & Signed progress \\
$\mathbf{a}^{\mathrm{op}}_t$ & Operator command &
$\Delta t^{\mathrm{w}}_k$ & Water time \\
$\mathbf{a}^{\mathrm{agent}}_t$ & Agent command &
$\pi_\theta$ & Policy \\
$\mathbf{a}_t$ & Applied command &
$V_\phi$ & Value function \\
$\mathcal{A}$ & Command space &
$\theta,\phi$ & Network parameters \\
$\beta$ & Assistance weight &
$\gamma$ & Discount factor \\
$\tau_t$ & Command age &
$\epsilon$ & PPO clip range \\
$\mathbf{o}_t$ & Observation &
$c_{\mathrm{ent}}$ & Entropy coefficient \\
$v_x,v_y$ & Velocity &
$\mathbf{s}_t$ & Simulation state \\
$\mathbf{f}=(f_x,f_y)$ & External force &
$\mathbf{h}_t$ & Haptic data \\
$\mathbf{f}_{\mathrm{d}},\mathbf{f}_{\mathrm{ch}}$ & Drag / restoring force &
$\hat{\mathbf{s}}$ & Cached state \\
$\mathbf{f}_c$ & Collision force &
$\hat{\mathbf{h}}$ & Cached haptic data \\
$F_{\max}$ & Max.\ restoring force &
 & \\
\bottomrule
\end{tabularx}
\end{table}

The course follows a water-free channel whose centre $y_c$ and height $w$ vary with the horizontal position and are defined by the course layout. Outside the channel lies the water zone, where the spaceship is slowed down and experiences a drag force $\mathbf{f}_{\mathrm{d}}$ opposing its motion. In addition, a vertical restoring force guides the spaceship towards the channel centre:
\begin{equation}
\begin{aligned}
\mathbf{f}_{\mathrm{ch}} &= -F_{\max}\,\rho^{2}\,\operatorname{sgn}(y_{\mathrm{ship}}-y_c)\,\hat{\mathbf{e}}_y,\\
\rho &= \min\!\left(\frac{2\,|y_{\mathrm{ship}}-y_c|}{w},\,1\right),
\end{aligned}
\label{eq:channel-force}
\end{equation}
where $y_{\mathrm{ship}}$ is the vertical position of the spaceship, $\hat{\mathbf{e}}_y$ is the vertical unit vector, and $\rho\in[0,1]$ is the normalised distance from the channel centre. The restoring force acts as a virtual fixture with maximum magnitude $F_{\max}=2$~N, reached at the channel boundary and maintained throughout the water zone. The quadratic profile provides gentle guidance near the centre and stronger resistance towards the boundaries. Because the water zone is invisible, the human operator relies on these forces to stay within the channel. Together with the contact force $\mathbf{f}_c$ arising from collisions with obstacles and boundaries, these forces form the two-dimensional external force
\begin{equation}
\mathbf{f} = \mathbf{f}_{\mathrm{d}} + \mathbf{f}_{\mathrm{ch}} + \mathbf{f}_c = (f_x, f_y),
\label{eq:external-force}
\end{equation}
which is rendered to the human operator as haptic feedback along the two in-plane axes of the device and used in the policy observation and reward.

\subsection{Agent Placement and Operator-Side Rendering}
\label{sec:architecture}

The trained policy is embedded in the same application as the remote virtual environment. Policy observations and actions are therefore exchanged within the application, giving the assistant direct access to the current simulation state without any network exchange. In contrast, human operator commands and state updates traverse the 5G network over the uplink and downlink, respectively. The application itself is hosted on an edge node co-located with the 5G core (Section~\ref{sec:edge}), which keeps the network path between the operator and the remote environment short.

At the human operator station, a replica of the environment is continuously corrected towards the state received from the remote environment and used for visual rendering, while the most recent received force is smoothed and held by the haptic device between updates. Unlike model-mediated teleoperation~\cite{mitra2008model}, haptic feedback is not computed from this replica. Network delay therefore affects the freshness of the state and force presented to the human operator, rather than the rate at which feedback is rendered.
\subsection{Assistive Agent}
\label{sec:agent}

The assistive policy is pretrained offline and deployed with fixed parameters, so the same policy is used across all network conditions.

\subsubsection{Observations and Shared Control}

The policy receives the four-dimensional haptic observation
\begin{equation}
\mathbf{o}_t = (v_x, v_y, f_x, f_y),
\label{eq:observation}
\end{equation}
comprising the spaceship velocity and the external force $\mathbf{f}$ from \eqref{eq:external-force}. The policy therefore relies on motion and force information without visual observations.

Let $\mathbf{p}^{\mathrm{op}}_t$ denote the measured position of the haptic device. A mapping $\mathcal{M}$ converts this position into the two-dimensional human operator command $\mathbf{a}^{\mathrm{op}}_t$:
\begin{equation}
\mathbf{a}^{\mathrm{op}}_t = \mathcal{M}(\mathbf{p}^{\mathrm{op}}_t).
\label{eq:operator-command}
\end{equation}
The command is transmitted to the remote endpoint, where the latest available value is used for shared control.

Both the human operator and the agent produce normalised two-dimensional velocity commands in $\mathcal{A}=[-1,1]^2$, which are scaled by the current speed limit to obtain the spaceship velocity. At each simulation step $t$, the two commands are combined through fixed-weight blending:
\begin{equation}
\mathbf{a}_t = (1-\beta)\,\mathbf{a}^{\mathrm{op}}_{t-\tau_t} + \beta\,\mathbf{a}^{\mathrm{agent}}_t,
\qquad
\mathbf{a}^{\mathrm{agent}}_t \sim \pi_\theta(\cdot \mid \mathbf{o}_t),
\label{eq:action-blending}
\end{equation}
where $\mathbf{a}^{\mathrm{op}}_{t-\tau_t}$ is the most recent human operator command available at remote environment, $\tau_t$ is its age, $\mathbf{a}^{\mathrm{agent}}_t$ is the agent command sampled from the policy given the current observation, and $\beta \in [0,1]$ is the assistance weight that determines the agent's share of the applied command $\mathbf{a}_t$. Because \eqref{eq:action-blending} is a convex combination, $\mathbf{a}_t \in \mathcal{A}$ for any $\beta$. The human operator command is thus subject to the network delay $\tau_t$, whereas the agent command is computed from the current state. Setting $\beta=0$ gives the human operator full control, whereas $\beta=1$ corresponds to full DRL-agent control. In the assisted condition, we set $\beta=0.5$, giving equal weight to both commands. This neutral setting avoids favouring either the human operator or the agent's policy and preserves substantial operator authority, while allowing the assistance to have a measurable effect on task execution. Keeping $\beta$ fixed also isolates the effect of the network conditions from that of the blending configuration.

The simulation advances with a fixed physics timestep $\delta t = 5$~ms, indexed by $t$. The DRL-trained policy acts at decision steps $k$, each spanning $D$ physics steps, so that the decision interval is $\Delta t_k = D\,\delta t$ and the most recent agent command is held between decisions. During deployment, the policy acts at every physics step ($D=1$), so the decision and physics steps coincide. A new agent command is therefore generated and blended with the latest available human operator command every 5~ms, i.e., at 200~Hz.

\subsubsection{Training Procedure}

The policy is trained without human-command blending or behavioural cloning. The training environment retains the water zone but excludes the gate obstacles. Training therefore focuses on navigation through the water zone, while the human operator subsequently controls the complete task with assistance from the trained policy. A new course is generated for each training episode.

We train the policy using Proximal Policy Optimization (PPO)~\cite{schulman2017ppo}, an actor--critic method that learns a policy $\pi_\theta$ and a value function $V_\phi$. The value function provides a baseline for estimating action advantages, and PPO updates the policy using a clipped probability-ratio objective that limits the size of each policy update. We select PPO for three reasons. First, it supports the continuous two-dimensional velocity actions required by the task. Second, its clipped updates provide stable learning with relatively little hyperparameter tuning, which is important for a policy that is later blended with human commands. Third, PPO is the standard, well-tested trainer in the Unity ML-Agents toolkit~\cite{juliani2018unity} used for the virtual game, allowing training to run directly in the same Unity environment that is deployed at the testbed.

\begin{table}[t]
\centering
\caption{Training Environment, Reward Weights, and PPO Hyperparameters}
\label{tab:training}
\resizebox{\columnwidth}{!}{%
\begin{tabular}{@{}ll@{}}
\toprule
\textbf{Parameter} & \textbf{Value} \\
\midrule
\multicolumn{2}{@{}l}{\textit{Environment}} \\
Course length & 500 units \\
Speed limit (air / water) & 30 / 8 units/s \\
Physics timestep $\delta t$ & 5~ms \\
Water exposure limit & 60~s \\
\midrule
\multicolumn{2}{@{}l}{\textit{Reward}} \\
Duration penalty $\kappa_{\mathrm{dur}}$ & 0.02~s$^{-1}$ \\
Progress reward $\kappa_{\mathrm{prog}}$ & 0.03~unit$^{-1}$ \\
Water-exposure penalty $\kappa_{\mathrm{w}}$ & 0.4~s$^{-1}$ \\
Vertical-force penalty $\kappa_y$ & 0.01 \\
Collision penalty $\kappa_c$ & 0.1 \\
Terminal rewards $R_{\mathrm{goal}}$ / $R_{\mathrm{fail}}$ & 5 / 2 \\
\midrule
\multicolumn{2}{@{}l}{\textit{Training (PPO)}} \\
Training steps $N$ & $3.6\times10^5$ \\
Decision period $D$ & 5 physics steps (25~ms) \\
Rollout buffer size $B$ & 8192 \\
Batch size & 256 \\
Learning rate & $3\times10^{-4}$ (constant) \\
Discount factor $\gamma$ & 0.99 \\
Clip range $\epsilon$ & 0.2 \\
Entropy coefficient $c_{\mathrm{ent}}$ & $5\times10^{-4}$ \\
\bottomrule
\end{tabular}%
}
\end{table}
\begin{figure*}[ht]
\centering
\includegraphics[width=\textwidth]{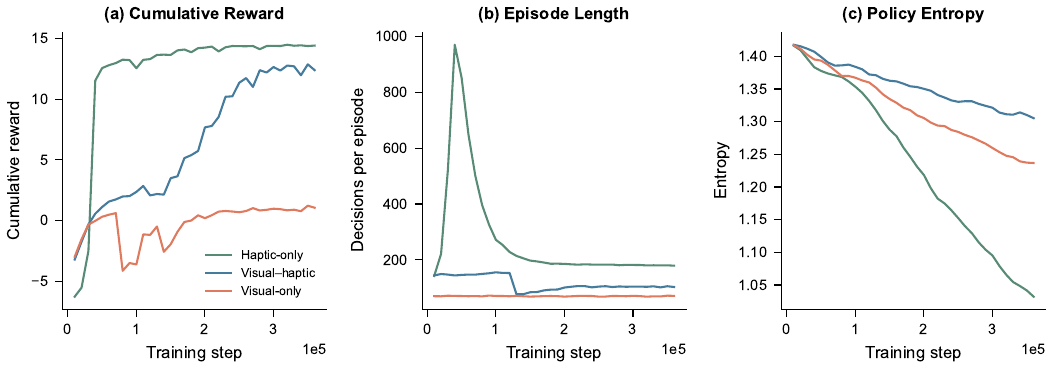}
\vspace{-0.3cm}
\caption{PPO training progression over $3.6 \times 10^5$ steps using haptic observations (velocity and external force components), visual observations (Unity camera images), or both: (a) cumulative reward, (b) episode length in policy decisions, and (c) policy entropy.}
\label{fig:training-curves}
\end{figure*}

The reward is assigned at each decision step $k$. Let $\Delta x_k$ denote the signed forward displacement and $\Delta t^{\mathrm{w}}_k$ the time spent in water zone during the decision interval $\Delta t_k$. The reward
\begin{equation}
\label{eq:reward}
\begin{aligned}
r_k ={}& -\kappa_{\mathrm{dur}}\,\Delta t_k + \kappa_{\mathrm{prog}}\,\Delta x_k - \kappa_{\mathrm{w}}\,\Delta t^{\mathrm{w}}_k \\
       & - \bigl(\kappa_y\,|f_y| + \kappa_c\,\|\mathbf{f}_c\|\bigr)\,\delta t + r^{\mathrm{term}}_k
\end{aligned}
\end{equation}
encourages forward progress while penalising elapsed time, water exposure, vertical force, and collisions. The force penalties are evaluated at the decision instant, with $f_y$ the vertical component of the external force in \eqref{eq:external-force}. An episode terminates when the spaceship crosses the finish line, accumulates 60~s of water exposure, or moves less than 5 units within a 5~s window; the last condition prevents prolonged episodes in which the agent remains stuck. The terminal reward is
\begin{equation}
\label{eq:terminal-reward}
r^{\mathrm{term}}_k =
\begin{cases}
+R_{\mathrm{goal}}, & \text{if the finish line is crossed},\\
-R_{\mathrm{fail}}, & \text{if the episode terminates otherwise},\\
0, & \text{at non-terminal steps}.
\end{cases}
\end{equation}
The weight values are listed in Table~\ref{tab:training}. The agent is trained to avoid the water zone over the fastest possible completion, so the learned policy prioritises leaving the water over speed, while the terminal rewards dominate the per-step terms and anchor the objective on finishing the course. The weights were selected empirically in preliminary runs and kept constant across all training configurations.

During training, $D=5$ ($\Delta t_k = 25$~ms), which reduces the number of policy evaluations per simulated second and accelerates training. Since the policy is memoryless, i.e., its action depends only on the current observation, the higher decision rate used during deployment leaves the learned mapping unchanged and only updates the agent command more often. Table~\ref{tab:training} summarises the training environment and PPO hyperparameters.

Fig.~\ref{fig:training-curves} compares three observation configurations over $3.6\times10^5$ training steps. The haptic-only policy receives the four velocity and force components used by the deployed assistant. The visual-only policy receives Unity camera images, while the visual--haptic policy combines these images with the same four scalar inputs.

The haptic-only policy improves fastest, exceeding a cumulative reward of 10 within approximately $4\times10^4$ steps and reaching 14.4 by the end of the plotted interval. The visual--haptic and visual-only policies reach approximately 12.4 and 1.0, respectively. For the haptic-only policy, episode length initially increases and subsequently decreases while reward remains high. Episode length reflects both navigation behaviour and the termination rules, hence shorter episodes alone do not establish better performance. Policy entropy decreases most strongly for the haptic-only policy, indicating a less dispersed action distribution. These results indicate that force and motion information provide a compact and informative observation for the assistance task, allowing the policy to learn faster and reach a higher reward than with visual input. The haptic-only policy is therefore used in the teleoperation experiments.

\section{5G Testbed and Measurement Setup}
\label{sec:testbed}

The assistive system is evaluated on a live standalone 5G testbed using production radio scheduling and QoS mechanisms. This section describes the network and edge deployment, the traffic conditions used to induce congestion, the QoS prioritisation mechanism, and the cross-layer instrumentation.

\subsection{KINET 5G Network}
\label{sec:network}

\begin{figure}[H]
    \centering
    \includegraphics[width=0.8\columnwidth]{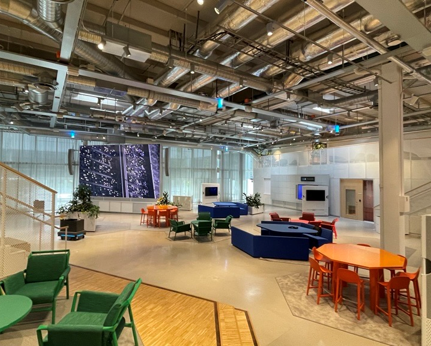}
    \caption{Indoor KINET testbed at Ericsson, showing the ceiling-mounted radio units (indicated by the blue lights) and the experimental area used for the teleoperation study.}
    \label{fig:kinet-photo}
\end{figure}

Experiments were conducted on the Ericsson KINET 5G standalone (SA) testbed at the Kista Innovation Park~\cite{testbed2026}. The human operator's client runs on a laptop connected to the network through a 5G USB modem. The modem is served by a single indoor Ericsson Radio Dot System (RDS) cell, which operates in the 3.5~GHz band (n78) with an 80~MHz time-division duplex (TDD) carrier and is connected to a shared 5G core network. Fig.~\ref{fig:kinet-photo} shows the indoor testbed, including the ceiling-mounted radio units serving the experimental area. RAN measurements collected by the testbed were visualised in real time using Grafana dashboards~\cite{grafana}.
Within each experimental block, reference signal received power (RSRP) varies by at most 6~dB, while the signal-to-interference-plus-noise ratio (SINR) remains above 33~dB. Radio measurements are reported in Section~\ref{sec:results} to characterise the conditions under which each comparison was performed.

\subsection{Edge Deployment}
\label{sec:edge}
The Unity application containing the virtual environment and the trained policy is packaged as a Docker container image, which bundles the application with all its dependencies so that it runs identically on any host. The container runs on an edge node co-located with the 5G core network and is managed by K3s, a lightweight distribution of the Kubernetes container-orchestration platform. The application is deployed with Helm, a package manager for Kubernetes, and the experiment parameters, such as the control mode, are set in its deployment configuration. Each experimental condition can therefore be selected without rebuilding the application image.

\subsection{Traffic Conditions}
\label{sec:traffic}

Congestion is introduced by generating competing downlink traffic to seven background user equipments (UEs), i.e., 5G devices sharing the same cell with the teleoperation UE. The background UEs exhibit measured downlink throughput of approximately 115--125\,Mbps each, driving cell utilisation close to saturation. The background traffic was started before the teleoperation application and ran continuously throughout each congested and prioritised block.

In addition to the background traffic, a synthetic video-like stream is transmitted to the operator UE. This allows us to evaluate haptic communication under both cell-level congestion and additional traffic sharing the operator UE's downlink. Although the demonstrator renders visual feedback at the operator station, the synthetic stream represents the communication load that remote video delivery could introduce in a teleoperation application. It is generated by a custom source that transmits from the server to the operator over the User Datagram Protocol (UDP). The stream sends frame bursts at 30 frames/s with an average bitrate of 100\,Mbps, using a separate socket from the teleoperation traffic. To mimic compressed video, every 30th frame is four times larger than the others. Each packet carries sequence numbers and a timestamp, allowing the receiver to measure throughput, packet loss, and jitter.

We evaluate three network conditions:

\begin{itemize}
\item \textbf{Uncongested:} Both the background UE traffic and the synthetic video-like stream are disabled. Mean downlink physical resource block (PRB) utilisation is 17.8\%.

\item \textbf{Congested:} Background traffic to seven competing UEs and the synthetic stream to the operator UE are enabled, increasing mean downlink PRB utilisation to 92.7\%.

\item \textbf{Prioritised:} The same traffic configuration is retained, while the teleoperation traffic receives QoS priority. Mean downlink PRB utilisation reaches 92.8\%.
\end{itemize}

Because the congested and prioritised conditions share the same competing traffic and closely matched PRB utilisation, differences in latency between them can be attributed to the QoS scheduling treatment rather than to differences in offered radio load.

\subsection{QoS Prioritisation Mechanism}
\label{sec:qos-mechanism}

Per-flow QoS prioritisation is requested through the CAMARA Quality-on-Demand (QoD) application programming interface (API)~\cite{camara_qod}. The requested QoS profile is mapped to 5G QoS Identifier (5QI)~8 in the testbed configuration. 5QI~8 is a non-guaranteed-bitrate class with a higher scheduling priority than the default 5QI~9 that carries the competing traffic.

The application container uses the host network namespace of the edge node, exposing the teleoperation uplink and downlink on UDP ports. The endpoint addresses, transport protocol, and port numbers identify the application flows for the per-flow QoS request, while the synthetic video-like stream uses a separate UDP socket, allowing it to be distinguished from the teleoperation traffic.

\subsection{Layered Instrumentation}
\label{sec:instrumentation}

Measurements are collected concurrently across the measurement layers listed in Table~\ref{tab:measurement-planes}. Each layer exports its metrics to a single, centrally hosted time-series database on the edge cluster, where they are stored under a common time base, so that UE, RAN, transport, and application samples from the same session can be aligned and analysed jointly.

\begin{table}[t]
\centering
\caption{Concurrent measurement layers.}
\label{tab:measurement-planes}
\footnotesize
\resizebox{\columnwidth}{!}{%
\begin{tabular}{@{}llll@{}}
\toprule
Layer & Source & Granularity & Principal quantities \\
\midrule
UE          & 5G USB modem & $\approx 3.4$~s  & RSRP, SINR, band, signal level \\
RAN         & Baseband     & 1~s              & PRB utilisation, CQI, RLC delay, throughput \\
Transport   & TWAMP        & $\approx 100$~Hz & One-way and round-trip delay \\
Application & Task logs    & Per session      & Completion time, water-zone time \\
Telemetry   & Game state   & 6~s              & Position, blended command \\
Compute     & cAdvisor     & 15~s             & CPU, memory, container I/O \\
\bottomrule
\end{tabular}}
\end{table}

Transport delay is measured using active TWAMP probes between the UE and a reflector~\cite{rfc5357}. Five packet sizes are multiplexed (50, 100, 200, 400, and 800~bytes), allowing packet-size effects to be separated from queueing behaviour. The aggregate probe rate is approximately 100~packets/s, resulting in 357,670 probes and 356,801 round-trip samples across the experiment.

The testbed streams RAN CTR measurements at 1~s intervals, covering cell, UE, radio-bearer, and baseband scopes. These scopes populate different fields within the same counter stream and are separated before analysis. This streaming access is specific to the testbed configuration. Within the network, performance measurements are typically collected over a configured recording output period (ROP) and are available to the network operator rather than directly to the application.

Application telemetry is exported at the 6~s interval of the testbed's monitoring pipeline, which we did not modify. With sessions lasting 33--45~s on average, this yields several samples per session, which is sufficient to associate each session with the corresponding network state. The telemetry is not used for the reported outcomes: task metrics are taken from the per-session task logs, and latency is characterised using the TWAMP probes. Because the interval is much longer than the 5~ms control period, we do not draw conclusions about command smoothness or reaction time.

\section{Experimental Methodology}
\label{sec:method}

\subsection{Experimental Design}
\label{sec:design}

The experiment follows a single-operator, repeated-measures block design with two factors: \emph{control mode} (Human-only, DRL-Assist) and \emph{network condition} (Uncongested, Congested, Prioritised). Each of the six factor combinations is executed as one continuous block of consecutive sessions lasting approximately 10--12 minutes, as summarised in Table~\ref{tab:block-summary}. A block design was used because changing the network configuration was time-consuming.

\begin{table}[t]
\centering
\caption{Experimental blocks in execution order. Session counts refer to completed sessions retained for analysis.}
\label{tab:block-summary}
\small
\renewcommand{\arraystretch}{1.15}
\setlength{\tabcolsep}{3pt}
\begin{tabular*}{\columnwidth}
{@{\extracolsep{\fill}}cllrr@{}}
\toprule
\textbf{Block} & \textbf{Condition} & \textbf{Mode}
& \textbf{Time} & \textbf{Sessions} \\
\midrule
1 & Congested   & Human-only     & 10 mins & 14 \\
2 & Congested   & DRL-Assist & 11 mins & 18 \\
\addlinespace
3 & Prioritised & DRL-Assist & 10 mins & 18 \\
4 & Prioritised & Human-only     & 11 mins & 17 \\
\addlinespace
5 & Uncongested & Human-only     & 10 mins & 16 \\
6 & Uncongested & DRL-Assist & 12 mins & 21 \\
\midrule
\multicolumn{4}{@{}l}{\textbf{Total analysed sessions}}
& \textbf{104} \\
\bottomrule
\end{tabular*}
\end{table}

 The user interface (UI) indicated the active control mode (Human-only or DRL-Assist) and displayed the application-layer round-trip time measured through periodic ping messages. This reflects a practical deployment, in which a human operator would be aware of whether assistance is enabled and could monitor the connection quality. The protocol includes two features that strengthen the comparison between control modes. First, the block order was reversed in the prioritised condition, where the DRL-Assist block preceded the Human-only block. DRL-Assist outperformed Human-only regardless of block order, indicating that the benefit of assistance is not explained by practice alone. Second, all six blocks were recorded on the same day by a single trained human operator, keeping operator and environmental conditions consistent across all control modes and network conditions.

\subsection{Session Protocol}
\label{sec:protocol}

Each session begins with the spaceship at a fixed starting position. The human operator navigates the course using the haptic device described in Section~\ref{sec:task}. The two control modes correspond to $\beta=0$ (Human-only) and $\beta=0.5$ (DRL-Assist) in \eqref{eq:action-blending}. A session ends when the course is completed.

Algorithm~\ref{drl-teleoperation} summarises the two processes that run concurrently during a session. The operator side reads the haptic device, transmits commands when due, and renders visual and haptic feedback at the operator station from the most recently received state $\hat{\mathbf{s}}$ and haptic data $\hat{\mathbf{h}}$. The remote side advances the simulation every $\delta t = 5$~ms: it blends the latest available operator command with the agent command according to \eqref{eq:action-blending}, and returns the updated state and haptic data to the operator. The two sides communicate asynchronously, so command transmission, feedback transmission, visual rendering, and the 1~kHz haptic servo each run at their own rate, independently of the 200~Hz simulation.
\begin{algorithm}[t]
\caption{DRL-assisted teleoperation loop}
\label{drl-teleoperation}
\begin{algorithmic}[1]
\Require Policy $\pi_\theta$, assistance weight $\beta$, timestep $\delta t$
\State Initialise states, command buffers and $t \gets 0$

\Statex \textbf{Operator side}
\While{session is active}
    \State $\mathbf{a}^{\mathrm{op}} \gets \operatorname{ReadDevice}()$
    \If{command transmission is due}
        \State $\operatorname{Send}(\mathbf{a}^{\mathrm{op}})$
    \EndIf
    \If{new feedback is received}
        \State $(\hat{\mathbf{s}},\hat{\mathbf{h}}) \gets \operatorname{Receive}()$
            \Comment{Refresh cached copy}
    \EndIf
    \State $\operatorname{RenderVisual}(\hat{\mathbf{s}})$
    \State $\operatorname{RenderHaptic}(\hat{\mathbf{h}})$
\EndWhile

\Statex \textbf{Remote side (concurrently)}
\While{session is active}
    \State $\mathbf{a}^{\mathrm{op}}_{t-\tau_t} \gets \operatorname{LatestCommand}()$
        \Comment{Delayed by network}
    \State $\mathbf{o}_t \gets$ observation of $\mathbf{s}_t$
        \Comment{Eq.~\ref{eq:observation}, current state}
    \State $\mathbf{a}^{\mathrm{agent}}_t \sim \pi_\theta(\cdot \mid \mathbf{o}_t)$
    \State $\mathbf{a}_t \gets (1-\beta)\,\mathbf{a}^{\mathrm{op}}_{t-\tau_t} + \beta\,\mathbf{a}^{\mathrm{agent}}_t$
        \Comment{Eq.~\ref{eq:action-blending}}
    \State $(\mathbf{s}_{t+1},\mathbf{h}_{t+1}) \gets \operatorname{Step}(\mathbf{s}_t,\mathbf{a}_t,\delta t)$
        \Comment{Advance simulation}
    \If{feedback transmission is due}
        \State $\operatorname{Send}(\mathbf{s}_{t+1},\mathbf{h}_{t+1})$
    \EndIf
    \If{$\mathbf{s}_{t+1}$ is terminal}
        \State End session
    \EndIf
    \State $t \gets t+1$
\EndWhile
\end{algorithmic}
\end{algorithm}

Before each block, the human operator performed a single practice run to become familiar with the active control mode and network condition, reducing learning effects within the recorded sessions. Sessions affected by equipment malfunction or network misconfiguration were excluded from the analysis. After filtering, the dataset contains \num{104} completed sessions: \num{47} Human-only and \num{57} DRL-Assist, collected across the six experimental blocks listed in Table~\ref{tab:block-summary}.

\subsection{Evaluation Metrics and Data Processing}
\label{sec:eval-metrics}

\subsubsection{Task-level metrics}

Task performance is evaluated using three session-level metrics:
\begin{itemize}
    \item \textbf{Completion time}: elapsed time from session start to course completion, in seconds.
    \item \textbf{Water-zone time}: cumulative time spent in the water zone, in seconds.
    \item \textbf{Water exposure ratio}: water zone time divided by completion time.
\end{itemize}

\subsubsection{Network-level metrics}

Network performance is characterised using the measurement layers summarised in Table~\ref{tab:measurement-planes} and described in Section~\ref{sec:instrumentation}. Transport-layer delay is given by the TWAMP round-trip time. At the application layer, each session log records the mean and maximum round-trip time, measured through periodic ping messages, together with the jitter, packet loss, and throughput of the synthetic video-like stream. The two delay measures are reported separately because they cover different measurement paths and processing stages. Radio and RAN measurements characterise cell load and radio conditions; they support the interpretation of the results but are not treated as outcome variables.

\subsubsection{Statistical analysis}

Task performance is compared between control modes across the three network conditions using session-level summary statistics. Within-session latency accumulation is characterised by fitting a linear regression of TWAMP round-trip time against elapsed session time. Session-boundary changes are summarised by comparing the mean latency in the final 5~s of each session with the first 5~s of the following session. The results describe a feasibility demonstration conducted within a single experimental campaign.
\section{Experimental Results}
\label{sec:results}

We first examine how congestion and QoS prioritisation affect network latency and task performance. We then compare human-only and DRL-assisted operation under each network condition.

\subsection{Impact of Network Conditions}
\label{sec:res-congestion}

\subsubsection{Network-Level Effects}

\begin{table}[t]
\centering
\caption{Radio conditions and cell utilisation. Values are means pooled over Human-only and DRL-Assist recordings.}
\label{tab:ran-measurements}
\small
\setlength{\tabcolsep}{3pt}
\renewcommand{\arraystretch}{1.15}
\begin{tabularx}{\columnwidth}{@{}l *{3}{>{\centering\arraybackslash}X}@{}}
\toprule
\textbf{Metric} &
{\textbf{\shortstack{Uncongested}}} &
{\textbf{Congested}} &
{\textbf{Prioritised}} \\
\midrule
RSRP (dBm)         & $-77.0$ & $-78.8$ & $-84.0$ \\
SINR (dB)          & 35.9    & 34.6    & 33.8    \\
CQI                & 12.7    & 11.5    & 11.3    \\
DL PRB usage (\%)  & 17.8    & 92.7    & 92.8    \\
UL PRB usage (\%)  & 15.0    & 12.6    & 12.8    \\
\bottomrule
\end{tabularx}
\end{table}
Table~\ref{tab:ran-measurements} summarises mean radio measurements and cell utilisation for each network condition, pooled across both control modes. Radio conditions were stable throughout: SINR remained between 33.5 and 35.9~dB in all blocks. RSRP was constant in four blocks and varied by at most 6~dB (between $-84$ and $-78$~dBm) in the Congested Human-only and Prioritised DRL-Assist blocks.

\begin{table}[t]
\centering
\caption{TWAMP round-trip time distribution by network condition (ms).}
\label{tab:twamp}
\footnotesize
\resizebox{\columnwidth}{!}{%
\begin{tabular}{@{}lrrrrrr@{}}
\toprule
Condition & Packets & Median & p95 & p99 & Max & IQR \\
\midrule
Uncongested & 119,295 & 14.8 & 22.2 & 29.7 & 90.7 & 4.8 \\
Congested   & 118,843 & 113.2 & 494.8 & 520.8 & 588.6 & 242.6 \\
Prioritised & 118,663 & 15.7 & 21.8 & 24.8 & 67.6 & 4.6 \\
\addlinespace
\textit{Cong.\,/\,Prio.} & --- & $7\times$ & $23\times$ & $21\times$ & $9\times$ & $53\times$ \\
\bottomrule
\end{tabular}}
\end{table}
\begin{figure}[!ht]
\centering
\includegraphics[width=\columnwidth]{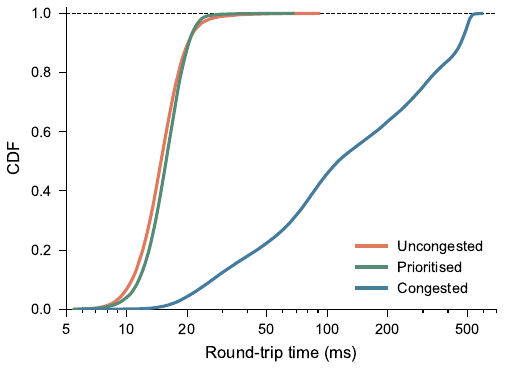}
\caption{CDF of per-packet TWAMP round-trip time by network condition.
Prioritised traffic recovers the uncongested baseline despite matched
radio load (\SI{92.8}{\percent} vs \SI{92.7}{\percent} DL PRB utilisation).}
\label{fig:latency-cdf}
\end{figure}
Congestion affects both the typical latency and its variability.
Table~\ref{tab:twamp} reports the TWAMP round-trip-time distribution under
each network condition. The median increases from 14.8~ms under the
Uncongested condition to 113.2~ms under the Congested condition. The 95th percentile
increases from 22.2~ms to 494.8~ms, while the interquartile range widens from
4.8~ms to 242.6~ms. Congestion therefore does not simply shift the latency
distribution; it also makes the delivered latency considerably less
consistent.

\begin{figure*}[ht]
\centering
\includegraphics[width=\textwidth]{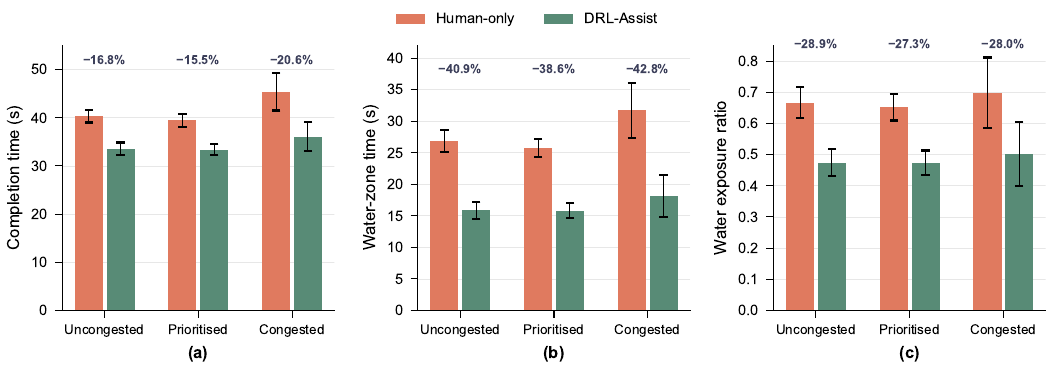}
\vspace{-0.3cm}
\caption{Task performance by network condition and control mode (mean $\pm$ SD).
(a)~Completion time, (b)~water-zone time, and (c)~water exposure ratio
(fraction of session time spent in the water zone). Percentage labels
indicate the reduction from Human-only to DRL-Assist.}
\label{fig:task-performance}
\vspace{-0.3cm}
\end{figure*}

With QoS prioritisation enabled, median RTT falls to 15.7~ms and the 95th percentile to 21.8~ms, approaching the Uncongested baseline despite similar mean downlink PRB utilisation in the Congested and Prioritised conditions (92.7\% and 92.8\%, respectively). Fig.~\ref{fig:latency-cdf} illustrates this recovery across the latency distribution, and further shows that only approximately 45\% of congested RTT samples fall below 100~ms.

\begin{figure}[!ht]
\centering
\includegraphics[width=0.95\columnwidth]{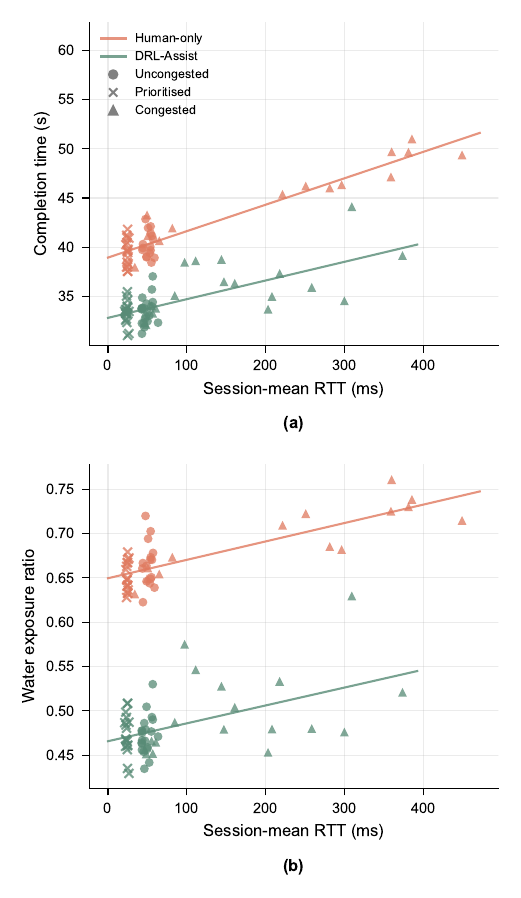}
\vspace{-0.5cm}
\caption{Task performance as a function of session-mean application-layer RTT. Each point represents one session, and the lines show separate linear regression fits for each control mode. Completion time increases more slowly with RTT under DRL-Assist, while the water exposure ratio is reduced by a similar margin across the RTT range.}
\label{fig:rtt-performance}
\vspace{-0.5cm}
\end{figure}
\subsubsection{Task-Level Effects}

The network degradation is also visible at the task level.
Figs.~\ref{fig:task-performance}(a) and (b)
report mean completion time and water-zone time, respectively, for each
control mode and network condition, with error bars indicating standard deviations. For Human-only control, mean completion
time increases from 40.31~s under the Uncongested condition to 45.42~s under the Congested condition. For DRL-Assist, it increases from 33.53~s to 36.07~s. Water-zone time follows the same pattern.

Congestion also makes task performance less consistent. The standard
deviation of completion time increases from 1.29~s to 3.92~s for Human-only
control and from 1.34~s to 2.99~s for DRL-Assist. This higher task-level
variability is consistent with the broader latency distribution observed at
the transport layer.

Under the Prioritised condition, mean completion time and water-zone time values are close to the Uncongested baseline for both control modes, consistent with the recovery observed in network latency.

\subsection{Impact of DRL Assistance}
\label{sec:res-assist}

As shown in Fig.~\ref{fig:task-performance}(a), DRL assistance reduces mean completion time under all three network conditions: by 6.8~s (16.8\%) under the Uncongested condition, 6.1~s (15.5\%) under the Prioritised condition, and 9.3~s (20.6\%) under the Congested condition. The largest observed reduction therefore occurs under the Congested condition. Water-zone time decreases by approximately 40\% in each condition, and the water exposure ratio by 27--29\%, as shown in Figs.~\ref{fig:task-performance}(b) and (c).

These results also clarify what QoS prioritisation can and cannot provide. Prioritisation removes the network-induced degradation: Human-only completion time under the Prioritised condition (39.48~s) returns to the Uncongested baseline (40.31~s). However, it cannot improve performance beyond what the operator achieves over an uncongested network, whereas DRL assistance reduces completion time further, to 33.37~s, under the same Prioritised conditions. Moreover, DRL-Assist under the Congested condition (36.07~s) outperforms Human-only under the Prioritised condition (39.48~s), and its water-zone time is lower as well (18.1~s versus 25.7~s). QoS prioritisation therefore restores, but cannot improve upon, the operator's uncongested performance, and it relies on the network granting prioritised treatment. DRL assistance improves task performance in every condition, including when no prioritisation is available.

Fig.~\ref{fig:rtt-performance} relates session-mean application-layer RTT to completion time in (a) and water exposure ratio in (b). Each point represents one session. Both task metrics worsen as RTT increases under both control modes. For completion time, the DRL-Assist fit is flatter than the Human-only fit (1.9~s versus 2.7~s per 100~ms of RTT), so the gap between the two modes widens as latency grows. For the water exposure ratio, the two fits have similar slopes, and DRL-Assist remains approximately 0.18 lower across the observed RTT range. DRL assistance therefore makes completion time less sensitive to high latency, while reducing water exposure by a consistent margin.

\section{Discussion}
\label{sec:discussion}

\subsection{Latency Accumulation}
\label{sec:accumulation}

Under the Congested condition, latency does not remain at a fixed elevated level. It continues to increase while a session is running. A linear regression of TWAMP round-trip time against elapsed session time gives a slope of $+5.5$~ms/s under the Congested condition. The corresponding slopes are $+0.007$~ms/s under the Uncongested condition and $+0.001$~ms/s under the Prioritised condition. Both are negligible compared with the Congested condition.

The mean round-trip time during the first 10~s of a Congested session is 118.0~ms. During the final 10~s, it reaches 164.3~ms, giving an average increase of 46.3~ms within a session.

\begin{figure}[t]
\centering
\includegraphics[width=\columnwidth]{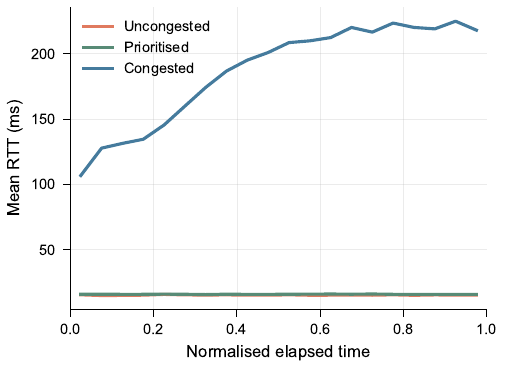}
\vspace{-0.3cm}
\caption{Mean TWAMP round-trip time over normalised session time. Elapsed
times between 5 and 25~s are mapped to $[0,1]$. Under the Congested condition, RTT
approximately doubles, while the Uncongested and Prioritised conditions
remain nearly flat.}
\label{fig:latency-ramp}
\end{figure}

Fig.~\ref{fig:latency-ramp} shows how RTT changes over the part of the session shared by all trials. We analyse elapsed times between 5 and 25~s and map this interval to $[0,1]$. The first 5~s are excluded to remove the initial start-up period, while the 25~s limit prevents the later points from being based only on longer sessions. Under the Congested condition, mean RTT approximately doubles across the observation window, from 106.8~ms to 218.1~ms. The Uncongested and Prioritised conditions remain nearly flat at approximately 15~ms.

Latency then falls when one session ends and the next begins. Comparing the final 5~s of each Congested session with the first 5~s of the following session gives a mean reduction of 50.9~ms. A steady increase during the task followed by a sharp drop at the boundary is consistent with a queue that fills while the application is active and partially drains between sessions. Neither pattern appears under the Prioritised condition, where the mean boundary drop is 0.0~ms. Prioritisation therefore suppresses the accumulation experienced by the teleoperation traffic even though the cell remains equally loaded.

These results indicate a coupling between task duration and network state. Longer sessions remain exposed to the growing queue for more time and therefore experience higher mean latency. This additional latency may then make the task harder and extend the session further. Under the Congested condition, session duration and session-mean latency are positively correlated. Controlling for session duration reduces the latency difference between the two control modes under the Congested condition by 80\%, indicating that part of the observed network difference is linked to how long the task remains active.

DRL assistance and prioritisation act on different sides of this coupling. Assistance shortens the task and reduces the operator's exposure to the latency increase, while prioritisation prevents the teleoperation traffic from entering the queue. This is also consistent with DRL assistance providing its largest benefit under the Congested condition, where the accumulation mechanism is active.

The use of non-visual observations also motivates future deployment on physical platforms. Where policy inference runs on an edge server connected to the remote device, transmitting compact state and force measurements could reduce the communication demand relative to transmitting images for policy inference. This may be useful when communication capacity is constrained. The benefit depends on sensor and inference placement and is not directly evaluated in the present experiments.

\subsection{Limitations}
\label{sec:limitations}

The study has four main limitations. First, all sessions were performed by a single trained human operator, so the results describe this operator and should not be generalised to a wider population. Second, the blocks were not randomised; although the order was reversed in the Prioritised condition, order effects cannot be fully excluded. Third, the evaluation uses a fixed assistance weight of $\beta = 0.5$ in the assisted condition. The experiments establish performance at this operating point but do not determine the optimal allocation of control between the operator and the agent. This allocation is a separate research question: it depends on the task, the haptic device, the type of manipulation, the operator's skill, and the network conditions, and may need to adapt during operation. We therefore fix $\beta$ to isolate the effect of the network conditions and leave the allocation problem to future work. Finally, the scene was rendered at the operator station rather than streamed as video, so the communication advantage of haptic over visual observations remains a design argument, and application telemetry was sampled every 6~s, limiting the application-layer analysis to session-level outcomes. Future work will address these limitations through multiple operators, randomised blocks, adaptive blending, and a fully remote video-haptic setup.

\section{Conclusion}
\label{sec:conclusion}

We developed a haptic-aware DRL assistant for teleoperation and evaluated it over a live Ericsson 5G SA testbed under Uncongested, Congested, and Prioritised conditions. For the evaluated operator, DRL assistance reduces mean completion time by 15.5--20.6\% and water-zone time by approximately 40\% across all conditions, with the largest improvement under the Congested condition. At matched radio load, QoS prioritisation requested through the CAMARA QoD API reduces median round-trip time from 113.2~ms to 15.7~ms and the 95th percentile from 494.8~ms to 21.8~ms, recovering latency close to the Uncongested baseline.

Under the Congested condition, round-trip time increases by 5.5~ms/s during a session and falls by 50.9~ms at session boundaries, coupling task duration to the latency the operator experiences. The two approaches address this coupling from different sides: prioritisation prevents the teleoperation traffic from entering the queue, while DRL assistance shortens the task and thereby reduces the operator's exposure to it.

Overall, the results indicate that communication, compute, and application functions should be co-designed as a single coupled system rather than optimised in isolation. In this work, embedding the DRL assistant in the remote application gave it direct access to the current environment state, edge placement kept the network path short, and a network API allowed the application to request QoS treatment for its own traffic. Network APIs are therefore a promising enabler for such co-design, exposing network capabilities and context to applications and compute platforms. A natural next step is to use network context obtained through such interfaces to adapt the blending between the human operator and agent as conditions change. 

\par
{\renewcommand{\baselinestretch}{0.9}\footnotesize
\bibliographystyle{IEEEtran}
\bibliography{references}\par}
\vspace{-1cm}
\raggedbottom
\begin{IEEEbiography}[{\includegraphics[width=1in,height=1.25in,clip,keepaspectratio]{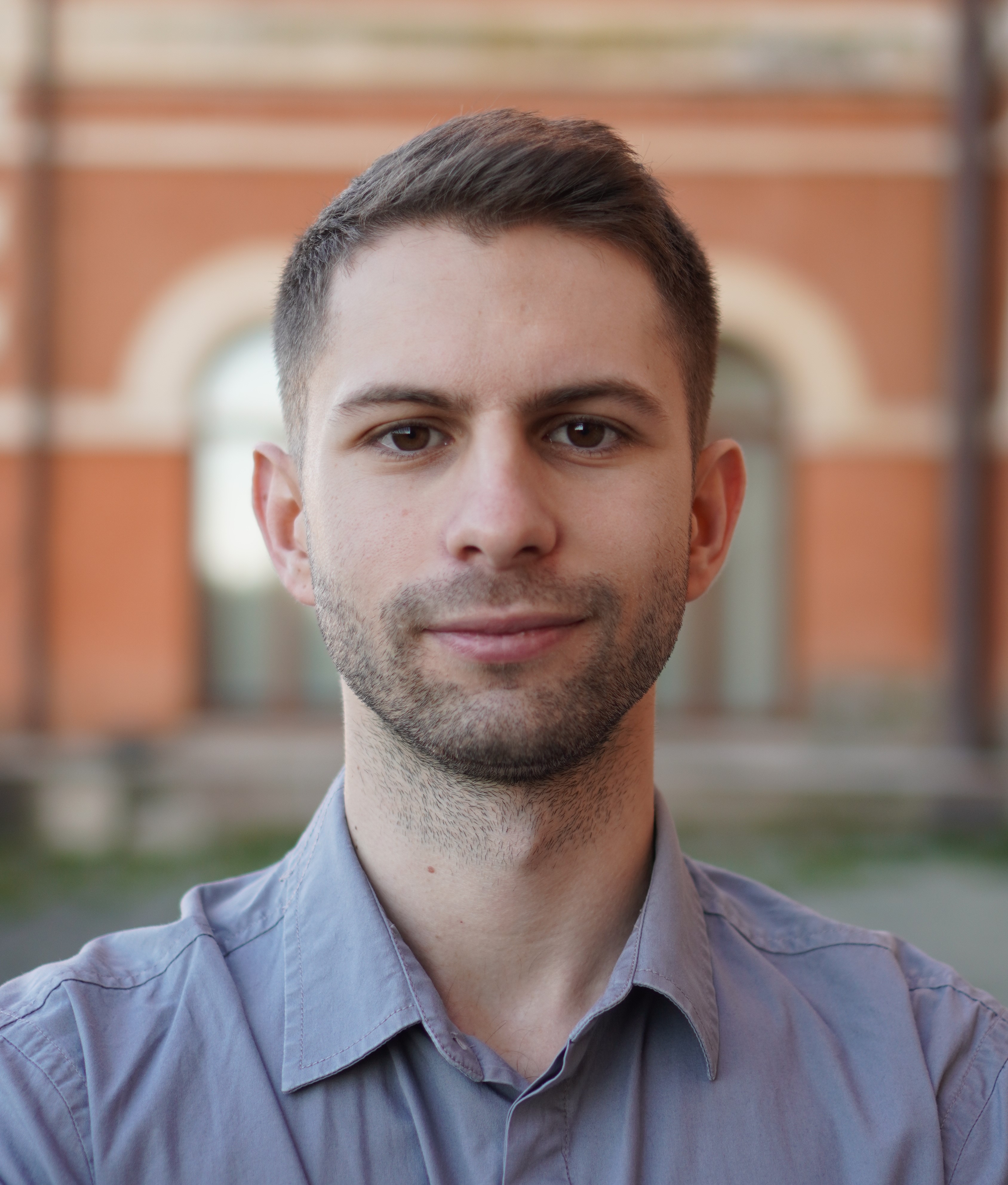}}]{Georgios Kokkinis (GS '25)}
is a Ph.D. candidate in the Department of Electrical and Computer Engineering at Aarhus University, Denmark, as a Doctoral Candidate of the TOAST MSCA Doctoral Network. Previously, he was a researcher with the Department of Electrical and Computer Engineering, University of Huddersfield, U.K. He received the Diploma (five-year, integrated M.Eng.) in Electrical and Computer Engineering from Aristotle University of Thessaloniki, Greece. His research interests include machine learning for wireless communications, the Tactile Internet and haptic communications, wireless networks, signal processing, and programming.
\end{IEEEbiography}

\vspace{-1cm}
\begin{IEEEbiography}[{\includegraphics[width=1in,height=1.25in,clip,keepaspectratio]{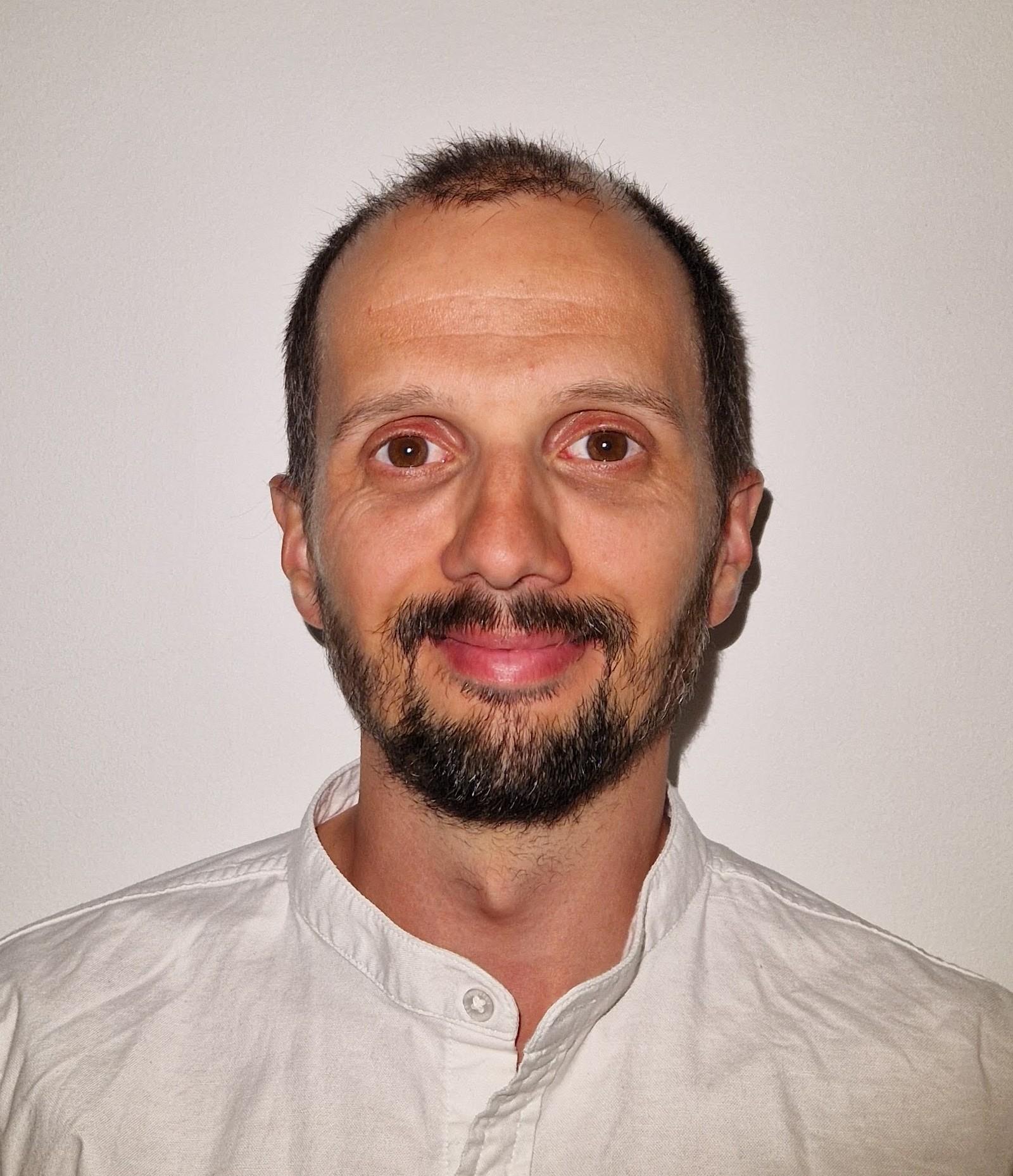}}]{Aitor Hernandez} 
is a Principal Researcher whose work focuses on cyber-physical systems, connected robotics, and physical AI. His research investigates how network APIs and service exposure can enable communication-compute-control co-design and unlock new capabilities for autonomous systems. He develops and evaluates real-world testbeds that integrate network intelligence into the cyber-physical loop. Aitor holds dual degrees in Telecommunications Engineering and Electronics Engineering from the Universitat Polit\`{e}cnica de Catalunya (UPC), Spain. \end{IEEEbiography}

\vspace{-1cm}

\begin{IEEEbiography}[{\includegraphics[width=1in,height=1.25in,clip,keepaspectratio]{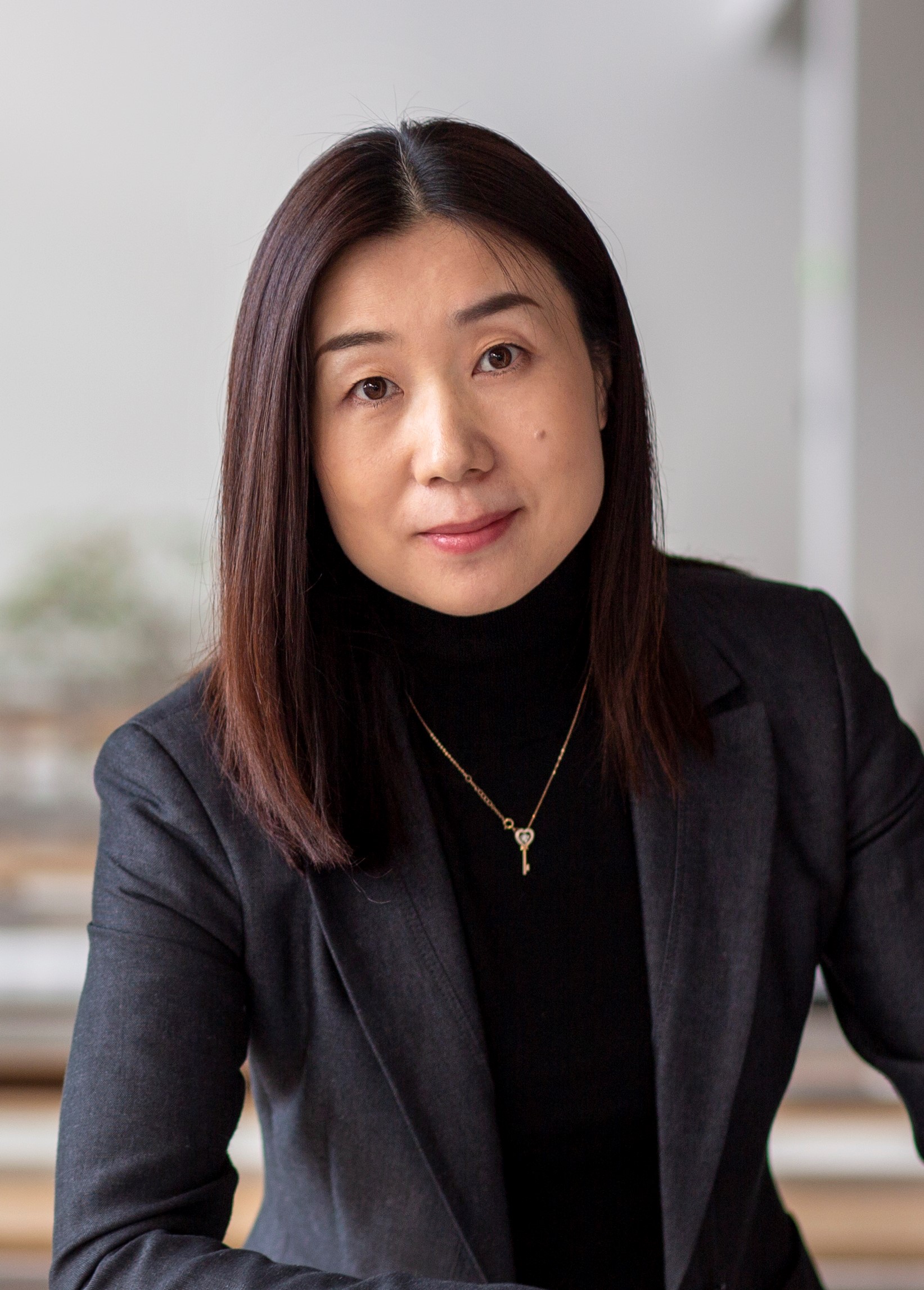}}]{Qi Zhang (SM '21)}
is a Professor with the Department of Electrical and Computer Engineering, Aarhus University, Aarhus, Denmark. She is leading the Internet of Things research area of AU Research Centre for Digitalisation, Big Data and Data Analytics (DIGIT). Her research interests include Internet of Things, Edge Intelligence, Tactile Internet, Goal-oriented Semantic Communication, as well as sensor data compression and analytics. She is the PI of three Danish Independent Research Fund Projects: AgilE-IoT (Agile Edge Intelligence for Delay Sensitive IoT) and Light-IoT (Analytics Straight on Compressed IoT Data) and eTouch (Edge Intelligence for Immersive Telerobotics in Touch-enabled Tactile Internet). She is the project coordinator and PI of Horizon Europe MSCA Doctoral Networks TOAST (Touch-enabled Tactile Internet Training Network and Open Source Testbed). She was an Associate Editor of EURASIP Journal on Wireless Communications and Networking. She has (co-)authored more than 150 publications in high impact journals and flagship conferences.
\end{IEEEbiography}

\vfill\null
\end{document}